\documentclass[aps,twocolumn,showpacs]{revtex4}
\usepackage{graphicx}
\usepackage{color}

\begin{document}

\title{Comparative study of vortex dynamics in CaKFe$_4$As$_4$ and Ba$_{0.6}$K$_{0.4}$Fe$_2$As$_2$ single crystals }

\author{Wang Cheng, Hai Lin, Bing Shen$^\dag$ and Hai-Hu Wen}\email{hhwen@nju.edu.cn}

\affiliation{Center for Superconducting Physics and Materials,
National Laboratory of Solid State Microstructures and Department
of Physics, National Center of Microstructures and Quantum
Manipulation, Nanjing University, Nanjing 210093, China}

\date{\today}

\begin{abstract}
We investigate the vortex dynamics in two typical hole doped iron based superconductors CaKFe$_4$As$_4$ (CaK1144) and Ba$_{0.6}$K$_{0.4}$Fe$_2$As$_2$ (BaK122) with similar superconducting transition temperatures. It is found that the magnetization hysteresis loop exhibits a clear second peak effect in BaK122 in wide temperature region while it is absent in CaK1144. However, a second peak effect of critical current density versus temperature is observed in CaK1144, which is however absent in BaK122. The different behaviors of second peak effect in BaK122 and CaK1144 may suggest distinct origins of vortex pinning in different systems. Magnetization and its relaxation have also been measured by using dynamical and conventional relaxation methods for both systems. Analysis and comparison of the two distinct systems show that the vortex pinning is stronger and the critical current density is higher in BaK122 system. It is found that the Maley's method can be used and thus the activation energy can be determined in BaK122 by using the time dependent magnetization in wide temperature region, but this is not applicable in CaK1144 systems. Finally we present the different regimes with distinct vortex dynamics in the field-temperature diagram for the two systems.
\end{abstract}

\maketitle

\section{Introduction}
After the discovery of superconductivity in LaFeAsO$_{1-x}$F$_x$\cite{HosonoJACS2008}, enormous studies have been conducted in iron-based pnictides or chalcogenides. They are regarded as the second family of unconventional high temperature superconducting systems. The study of vortex is very essential for iron based superconductors\cite{Toshihiro,Senatore,Prozorov,HuanYangAPL,BingShenPRB} since it is directly related to the high power applications. For a type-II superconductor, it is known that if the applied magnetic field is larger than the lower critical field $H_{c1}$, vortices will penetrate into the superconductor and the so-called mixed state is formed. The Lorentz force acting on vortices given by the external current $j_s$ will drive the vortices to move, and this motion is enhanced by the thermal fluctuations but hindered by the vortex pinning. Therefore the Lorentz force per unit length of vortex $f_L$ = $\Phi_0j_s$, the intrinsic pinning energy $U_c$ or activation energy $U(j)$, the elastic energy of vortex line or vortex bundle $E_{el}$ and the thermal energy $k_BT$ play complex roles in balancing the vortex motions, leading to very rich physics of vortex dynamics. Until now, there have been many theoretical models to explain the motion of vortices in superconductors, while the basic one is the so-called thermally activated flux motion model (TAFM)\cite{Anderson}. Concerning the activation energy for vortex motion, the collective pinning model\cite{Feigel} and vortex glass\cite{Fisher} theory have been widely used. Investigations on the vortex dynamics based on above theories were carried out in many iron-based superconductors\cite{HuanYangPRB,Haberkorn,YiYu,ShyamSundarPRB}.

Among the various phenomena of vortex dynamics, the second peak effect of magnetization is one of the most studied one, which seems to be quite common and shows up in many type-II superconductors. The basic feature is that the magnetization hysteresis loop exhibits a second peak when the magnetic field is increased except for the one near zero field. This phenomenon was observed in different families of iron based superconductors, including FeTe$_{x}$Se$_{1-x}$(11)\cite{Bonura,Das}, LiFeAs(111)\cite{Pramanik}, PrFeAsO$_{0.9}$\cite{VanderBeek} and SmOFeAsO$_{1-x}$F$_{x}$\cite{HuanYangPRB} (1111), Ba$_{1-x}$K$_x$Fe$_{2}$As$_{2}$ (122)\cite{Nakajima,Shegeyuki}. And there also exist various theoretical models or pictures to explain this phenomenon, including the transition between different vortex structures or regimes, a crossover from elastic to plastic vortex creep or order-disorder transition, etc. However, it is also found that the second peak effect may disappear in some systems, or for different doping levels in one system, such as in Co doped BaFe$_2$As$_2$\cite{BingShenPRB}. It is still unclear what is the definite reason for second peak effect, or whether there is a common reason for that in different systems.

Regarding the 122 family, the optimized $T_c\approx$ 40 K was obtained in Ba$_{0.6}$K$_{0.4}$Fe$_2$As$_2$ (BaK122) which is hole doped. Resistive measurements under magnetic fields show a very small anisotropy and extremely high upper critical field $H_{c2}$ and irreversibility field $H_{irr}$\cite{ZhaoshengWang}. Recently a new system CaKFe$_4$As$_4$ (CaK1144)\cite{Akiraiyo} was discovered with $T_c\approx$ 36 K. Experiments have shown that this system is also dominated by hole conduction. It is thus quite interesting to have a comparative investigation of vortex dynamics of these two systems. In this paper, we study the vortex dynamics for these two kinds of pnictide superconductors by measurements of magnetization and its relaxation. Our results reveal different features of vortex dynamics in these two systems.

\section{Experimental details}
The single crystals of CaK1144 and BaK122 were grown with the self-flux method. The growing processes of samples are the same as that reported elsewhere\cite{Akiraiyo,Meier,Luohuiqian}. The dimensions of two typical samples investigated here are 1.52*1.40*0.04 mm$^3$ (CaK1144) and 1.83*1.04*0.20 mm$^3$ (BaK122), respectively. In order to check the quality of the samples, we have done x-ray diffraction (XRD), resistivity and magnetization measurements.  The XRD measurements were performed on a Bruker D8 Advanced diffractometer with the Cu-K$_\alpha$ radiation. Only the sharp (00l) peaks were observed (not shown here), indicating the good crystallinity of the samples. The DC magnetization measurements were carried out with a SQUID-VSM-7T (Quantum Design). The resistive measurements were done with the four-probe method on a Quantum Design instrument Physical Property Measurement System (PPMS). The magnetization-hysterisis-loops (MHLs) for two samples were measured in magnetic fields up to 7 T. We also analyzed the vortex dynamics with the so-called dynamical\cite{Jirsa1997} and conventional\cite{Yeshurun} magnetization relaxation method. In the case of conventional relaxation measurement, the sample was zero-field cooled from above $T_c$ to the desired temperature. After that, the magnetic field was increased to a certain value and the time dependence of magnetization was measured immediately after stoping the field sweeping.

\section{Results and discussion}
\subsection{Magnetization and critical current density}

\begin{figure}
\includegraphics[width=8.5cm]{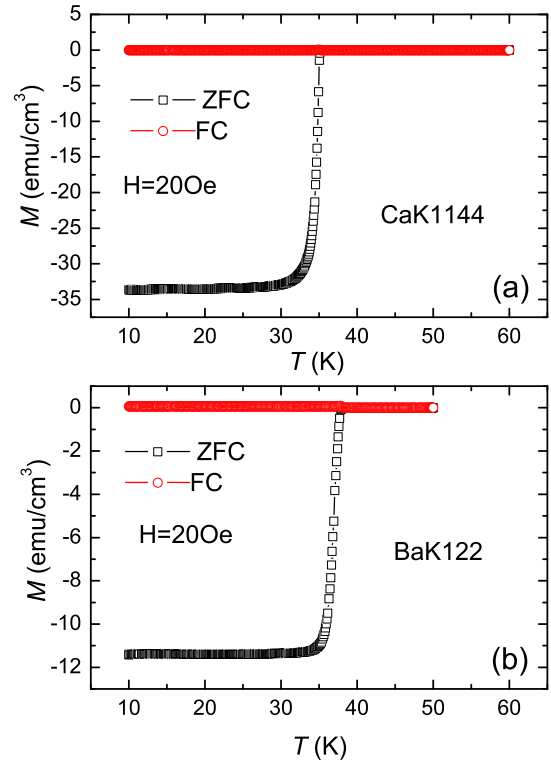}
\caption{(Color online) Temperature dependence of magnetization obtained with zero-field cooled and field cooled modes for (a)CaK1144 (b)BaK122 at $H$ = 20 Oe with $H\|c$.}\label{Fig1}
\end{figure}

 Fig.~\ref{Fig1} (a) and (b) show the temperature dependence of magnetization of these two samples after the process of zero-field cooled (ZFC) and field cooled (FC). The external applied magnetic field was 20 Oe and was parallel to c-axis of samples. From these figures, it is clearly seen that the superconducting transitions are very sharp for both samples. The superconducting transition temperatures determined here are about 35.0 K for CaK1144, and 37.8 K for BaK122. Therefore the $T_c$ values for both samples are quite close to each other. The sharp transition also indicates the good quality of our crystals. Estimate of magnetization in the ZFC mode tells a full magnetic shielding for both samples. The tiny positive magnetization signal in FC mode for CaK1144 may be due to the scan length problem of our SQUID in the iterating and fitting process. Since we mainly focus on the large magnetization measured in the ZFC mode, this will not give the problem for our subsequent data analysis.

\begin{figure}
\includegraphics[width=8.5cm]{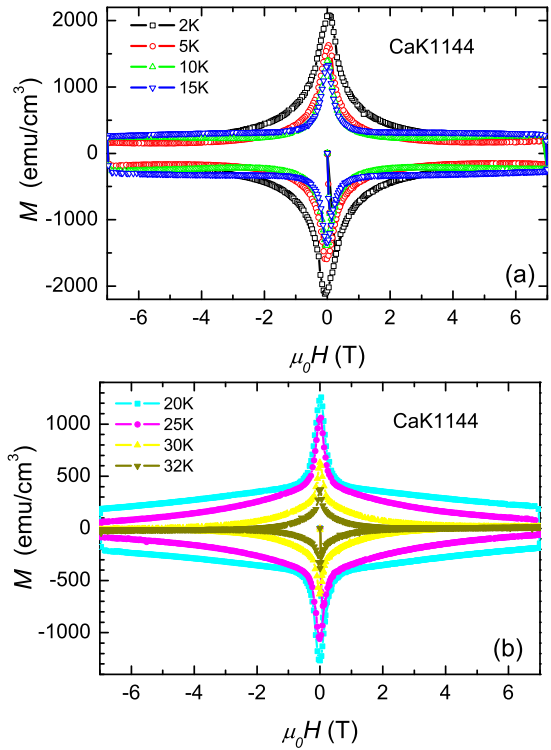}
\caption{(Color online) The isothermal MHLs of CaK1144 at (a)2, 5, 10, 15 K, and (b)20, 25, 30, 32 K.} \label{Fig2}
\end{figure}

Since the second peak effect is a common phenomenon for doped 122 family\cite{BingShenPRB,Shegeyuki,Salemsugui,Prozorovsecond}, it is curious to know whether it exists also for the CaK1144 system. In order to compare the difference of vortex dynamics between CaK1144 and BaK122, we measure the MHLs for two samples with a field sweeping rate of $dH/dt$ = 200 Oe/s. The results for CaK1144 are shown in Fig.~\ref{Fig2} (a) and (b). As usually observed in many superconductors, the width of MHLs increases with decreasing temperature due to more stronger pinning at low temperatures. Besides, the MHLs shown in Fig.~\ref{Fig2} (a) and (b) are very symmetric respective to the horizontal coordinate. It is known that the vortices can be pinned not only by the impurities or defects, but also the surface of the sample. According to the Bean critical state model\cite{Bean}, the former usually give rise to a symmetric MHL in the field increasing and decreasing processes. While the surface barrier or geometrical effect will lead to asymmetric MHLs\cite{YangRenSun}. Thus the MHLs measured here for CaK1144 show typical examples for the bulk pining. As one can see that a sharp peak of $M(H)$ occurs near zero magnetic field at all temperatures. The general reason for this magnetization peak near zero field is that, when the field is crossing zero from positive to negative, the entry and exit vortices will annihilate near the edge of sample, leading to an expediting escape of vortices from the interior and thus a high slope of $B(X)$. This gives rise to a large current density near the edge. In previous studies it was shown that, in the iron based superconductors, the magnetization peak near zero is much sharper than that in cuprates, which is attributed to the presence of strong pinning centers\cite{Yang,VanderBeek}. Very surprisingly, the MHLs in CaK1144 exhibit no clear second magnetization peak on most of the MHLs. This is unlike most of doped 122 systems\cite{BingShenPRB,Shegeyuki,Salemsugui,Prozorovsecond,Sun}. We will show that, instead of the absence of the second peak of magnetization versus magnetic field, a second peak effect shows up on the curve of MHL width or critical current density versus temperature. This phenomenon will be discussed in more details below.

\begin{figure}
\includegraphics[width=8.5cm]{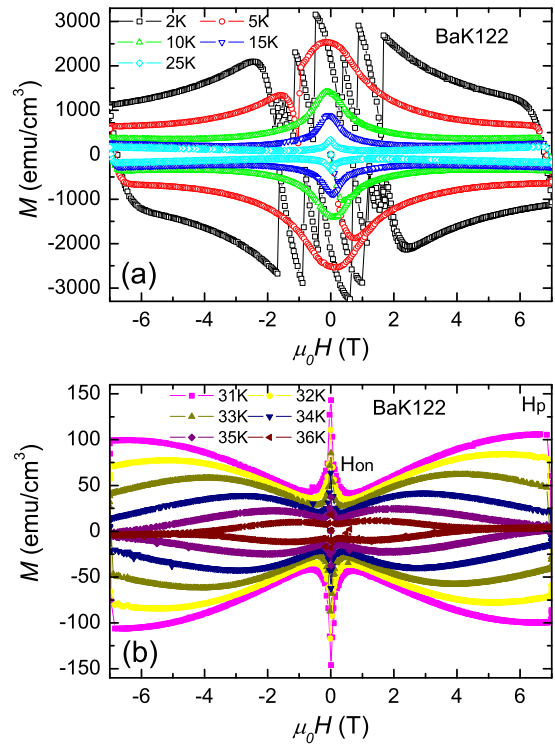}
\caption{(Color online) MHLs of BaK122 at various temperatures below $T_c$,(a)2, 5, 10, 15, 25 K,(b)15, 18, 20, 23, 25, 28 K,(c)31, 32, 33, 34, 35, 36 K.} \label{Fig3}
\end{figure}

Fig.~\ref{Fig3} (a) and (b) present the MHLs of BaK122 measured at several temperatures with a sweeping rate of $dH/dt$ = 200 Oe/s. Focusing on the MHLs at low temperatures in the range from 2 K to 5 K, we find that there are some big steps of magnetization, which are induced by the flux jumps for the BaK122 system. This is however absent in CaK1144 samples. Similar phenomenon is also observed in Ba(Fe$_{1-x}$Co$_x$)$_2$As$_2$ or BaNa122\cite{BingShenPRB,Pramaniksecond}. We must mention that in some cases, this flux jump is not observed in BaK122\cite{HuanYangAPL}. Fig.~\ref{Fig3} (b) presents the MHLs measured at temperatures close to $T_c$. A noticeable second peak effect is found on the MHLs in BaK122. From the minimum and maximum of magnetization we can determine the magnetic fields $H_{on}$ and $H_{p}$, as marked in Fig.~\ref{Fig3} (b) at different temperatures. As for lower temperatures, the peak appears at magnetic fields higher than the measurable field 7 T of our setup. The appearance of second peak effect in BaK122 and absence in CaK1144 indicate that the vortex dynamics are different in the two systems.

\begin{figure}
\includegraphics[width=8.5cm]{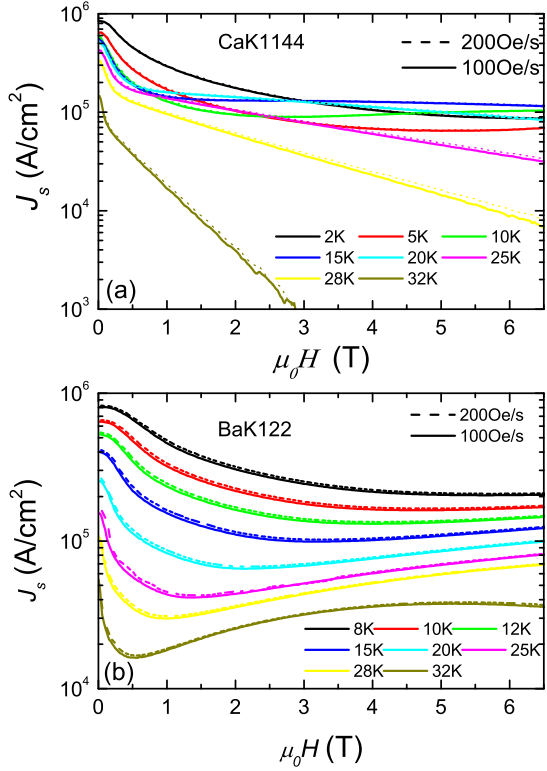}
\caption{(Color online) Correlations of $J_{s}$ versus $H$ for (a) CaK1144, and (b) BaK122 at different temperatures with $dH/dt$ = 200 Oe/s (dashed line) and 100 Oe/s (solid line). The second peak effect in BaK122 leads to a nonmonotonic changing of magnetization versus magnetic field.} \label{Fig4}
\end{figure}

From the MHLs plotted above, we can calculate the transient critical current density $J_{s}$ by using the Bean critical state model\cite{Bean} for both samples. The formula for the calculation reads
\begin{equation}\label{}
J_{s}=20\frac{\Delta M}{a(1-a/3b)},
\end{equation}

where $\Delta M=M_{+}-M_{-}$, $M_{+}$ and $M_{-}$ are the magnetization of the sample with decreasing and increasing magnetic field, $a$ and $b$ are the width and length of the sample ($b>a$). The results are shown in Fig.~\ref{Fig4} (a), (b) at different temperatures for CaK1144 and BaK122. The derived $J_{s}$ is quite high and reaches $8\times 10^{5}A/cm^{2}$ for CaK1144 at 2 K and 0 T, and for BaK122 at 8 K and 0 T. These values are comparable to most of doped samples in 122 family\cite{BingShenPRB,Shegeyuki,Yamamoto}. The $J_{s}(H)$ curves in Fig.~\ref{Fig4}(b) show the similar behavior of Na(Fe$_{1-x}$Co$_{x}$)As\cite{Ahmad} where $J_{s}(H)$ dependence can be divided into four regimes. In the low magnetic field regime, log$ J_{s}(H)$ reveals a small plateau. In the second regime, $J_{s}$ drops down with magnetic field and seems to satisfy a power law relationship $J_{s}\propto H^{-\alpha}$. As for the third regime, $J_{s}(H)$ exhibits an enhancement, which is found in previous studies on many other superconductors\cite{HuanYangAPL,BingShenPRB,Shegeyuki,KonczykowskiPRB2012}. However, in CaK1144 this second peak effect is absent or very weak. At higher temperatures, the third regime, or the so-called second-peak effect of $J_{s}(H)$ disappears. And compared to BaK122, $J_{s}$ drops more quickly for CaK1144 with increasing temperature and magnetic field. Because the maximum magnetic field of our SQUID is 7 T, the fourth regime of $J_{s}(H)$ with a decreasing behavior is not observed here for BaK122. From the variation of $J_{s}$ with magnetic field for both samples, we notice that the temperature and magnetic field play different roles in determining the vortex dynamics of the two systems. At low temperature and magnetic field, the vortex dynamics is within the scenario of single vortex pinning for both samples. At high temperature and magnetic field, vortex entanglement and plastic motion of vortex dislocations may occur leading to complicated features of magnetization for BaK122. For CaK1144, the vortex motion or pinning seem to change more strongly versus temperature, giving rise to a non-monotonic temperature dependence of $J_s(T)$ at a fixed magnetic field. From the absolute values of critical current density in the two systems at the same temperature and magnetic field, we may conclude that the vortex pinning in BaK122 is stronger than that in CaK1144.

\begin{figure}
\includegraphics[width=8.5cm]{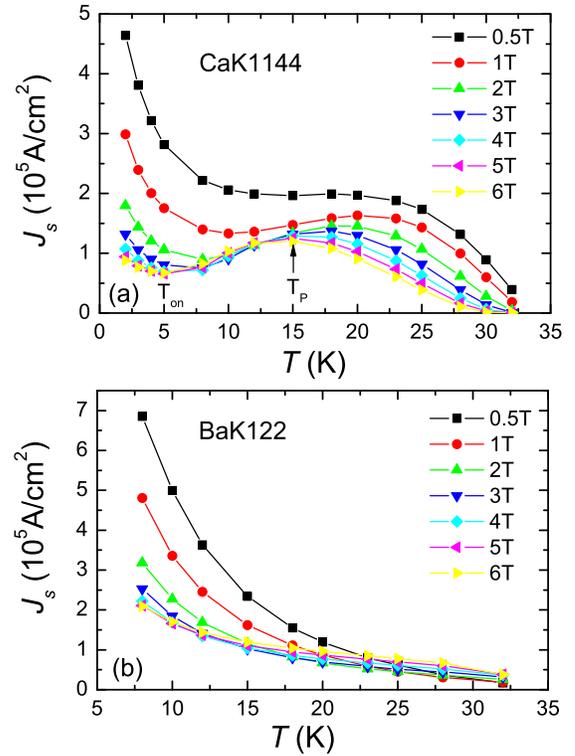}
\caption{(Color online) Temperature dependence of $J_s$ at 0.5, 1, 2, 3, 4, 5, 6, 7 T for (a) CaK1144, and (b) BaK122.} \label{Fig5}
\end{figure}

Fig.~\ref{Fig5}(a) and (b) present the temperature dependence of $J_{s}$ extracted from the MHLs at different magnetic fields for CaK1144 and BaK122, respectively. It is clear that the temperature dependence of $J_{s}$ in (a) and (b) are quite different. For BaK122, the temperature dependence of $J_{s}$ at different magnetic fields is monotonic. For CaK1144, however, the curve $J_{s}(T)$ initially decreases with increasing temperature and shows a minimum at temperature $T_{on}$, then increases and exhibits a peak at a certain temperature $T_{p}$. There exists a second peak effect of MHL width or critical current density $J_s$ versus temperature for CaK1144. With increasing the magnetic field, the temperatures associated with the peak and the minimum shift to lower values. The appearance of second peak effect of magnetization versus temperature suggests that the pinning force is enhanced notably at intermediate temperature region for CaK1144, unlike that in BaK122. It should be noticed that the peak effect of $J_{s}(T)$ for CaK1144 at 0.5 T is less clear, instead it shows a strong shoulder at higher temperatures. This is unlike the situation where the magnetic field is larger than 1 T. And we will give a qualitative explanation for the different "peak effect" for the two different systems in section D.

\subsection{Dynamical, conventional relaxation rate and activation energy}
Magnetization relaxation is an effective way to study the vortex dynamics in a superconductor. To study the vortex dynamics, dynamical and conventional magnetic relaxation are generally applied. The approach of dynamical magnetic relaxation is that the MHLs of a sample are measured with different sweeping rates of magnetic field at a fixed temperature. In this case one measures the correlation of the width of MHL and the magnetic field, namely $\Delta M$ vs $H$ with different sweeping rate $dB/dt$, and the dynamical relaxation rate is defined as
\begin{equation}\label{eqq}
  Q=\frac{d\ln(J_{s})}{d\ln(dB/dt)}=\frac{d\ln(\Delta M)}{d\ln(dB/dt)}.
\end{equation}

As for the conventional magnetic relaxation, a magnetic field is applied after the process of zero-field cooling at a constant temperature, then the time dependence of magnetization is measured immediately after stoping the sweeping of magnetic field. The magnetization relaxation rate is determined through

\begin{equation}\label{eqs}
 S=-\frac{d\ln(|M|)}{d\ln t}.
\end{equation}
Although these two relaxation rates are obtained by using two distinct ways, it has been shown that to some extent, $Q$ and $S$ are equal to each other and provide similar information about the vortex dynamics\cite{Vandalen,YanjingJiao}.
To analyze the experimental data, we adopt the model of Anderson\cite{Anderson} for thermally activated vortex motion, namely,

\begin{equation}\label{eqelectric}
 E=v_{0}B_e exp(\frac{-U(J_{s},T,B_{e})}{k_{B}T}).
 \end{equation}

 Here $E$ is the electric field caused by vortex motion, $v_{0}$ is the attempting moving velocity, $U$ is the activation energy which is also called the barrier energy depending on the transient current density $J_{s}$, temperature $T$ and the external magnetic field $B_{e}$. Furthermore, activation energy $U$ is generally expressed as\cite{Malozemoff},
\begin{equation}\label{equ}
  U(J_{s},T,B_{e})=\frac{U_{0}(T,B_{e})}{\mu (T,B_{e})}[(\frac{J_{c}(T,B_{e})}{J_{s}(T,B_{e})})^{\mu (T,B_{e})}-1].
\end{equation}

Where $U_{0}$ is the intrinsic pinning energy in the absence of driving force. The glassy exponent $\mu$ can take different values for different vortex pinning models or regimes.  For example, $\mu$ = -1 is corresponding to the linear relationship $U(J_s)=U_c(1-J_{s}/J_{c})$ or called as Kim-Anderson model\cite{Anderson}, while $\mu$ = 0 corresponds to the logarithmic model\cite{Zeldov}. For collective pinning or vortex glass model, the exponent $\mu$ generally takes a positive value between 0 and 2\cite{Feigel,Fisher}. Based on these basic functions and the definition of $Q$ and $S$, finally the following expressions are derived\cite{WenHH1995}.

\begin{equation}\label{eqUQ}
 \frac{T}{Q(T,B_{e})}=\frac{U_{0}(T,B_{e})}{k_{B}}+\mu(T,B_{e})CT.
\end{equation}
\begin{equation}\label{eqUS}
S=\frac{k_{B}T}{U_{0}+\mu k_{B}T\ln(t/t_{0})}.
\end{equation}
 Here $t_{0}$ is the characteristic relaxation time. By analyzing the data of $Q$ and $S$ under different conditions, we can get the information about the vortex motion in a superconductor.

\begin{figure}
\includegraphics[width=8.5cm]{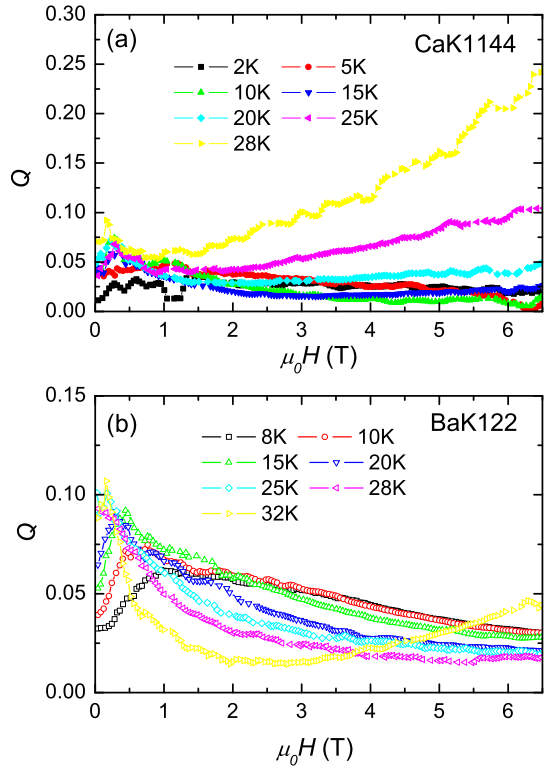}
\caption{(Color online) Dynamical magnetization relaxation rate as a function of magnetic field for (a)CaK1144, at 2, 5, 10, 15, 20, 25, 28 K (b)BaK122, at 8, 10, 15, 20, 25, 28, 32 K.} \label{Fig6}
\end{figure}

Fig.~\ref{Fig6}(a) and (b) show dynamical relaxation rate as a function of magnetic field for CaK1144 and BaK122, respectively. The parameter $Q$ is shown in Fig.~\ref{Fig6} with the calculation by Eq.~\ref{eqq} in which $J_{s}$ is measured with two different field sweeping rates 100 and 200 Oe/s. For CaK1144, the $Q(H)$ curve shows a small peak when the magnetic field is lower than 1 T.  In addition, the dynamical relaxation rate $Q$ changes slightly with magnetic field at low temperatures, generally in the region of about 5$\%$ or less for both samples. However, it increases obviously with magnetic field at higher temperatures, for example, at 25 K and 28 K for CaK1144. This suggests that thermal activation plays the dominant role in vortex motion at high temperatures. For BaK122, as shown in Fig.~\ref{Fig6}(b), there is also a peak of $Q$ in the low field region. With increasing temperature, this peak of $Q$ shifts to lower fields. Below the magnetic field corresponding to this peak, the MHL and $J_s$ both exhibit strong peaks, indicating a strong pinning of vortices in this region\cite{VanderBeek}. The value of $Q$ is in a small range without obvious increase when the magnetic field is larger than 1 T. Furthermore, $Q$ is less than 0.1 even at 32 K which is close to the superconducting transition temperature. This indicates the stronger vortex pinning at high temperatures in BaK122, compared to CaK1144.

\begin{figure}
\includegraphics[width=8.5cm]{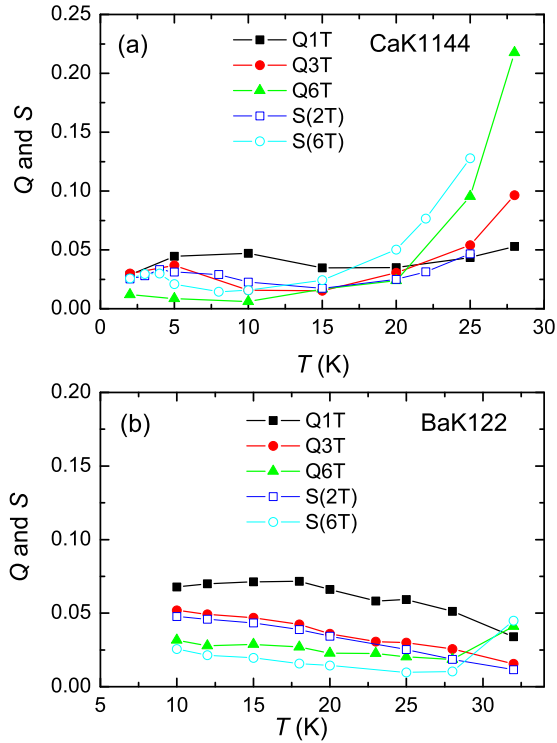}
\caption{(Color online) Temperature dependence of relaxation rate $Q$ and $S$ for (a) CaK1144, (b) BaK122 at different magnetic fields ranging from 1 to 6 T.} \label{Fig7}
\end{figure}

In Fig.~\ref{Fig7}(a) and (b), we present the temperature dependence of relaxation rate $Q$ and $S$ at different magnetic fields for CaK1144 and BaK122, respectively. At first, we discuss the dynamical magnetization relaxation rate $Q$. In Fig.7(a) for CaK1144, one can see that Q keeps a small value at low temperatures, but increases drastically when the temperature is higher than 20 K. One of the explanations for this rapid increasing of $Q$ value, as described in BaFe$_{2-x}$Co$_x$As$_2$\cite{BingShenPRB}, is that the plastic vortex creep is dominating in this regime rather than the elastic vortex creep. It is the dislocation of vortex that moves very fast leading to a rapid increasing of relaxation rate. When focusing on data in the intermediate temperatures region, say from 10 K to 20 K, we find that there exists a minimum value on each curve $Q(T)$ at magnetic fields above 2 T. This results in a concave shape for the plotted curve. A closer look finds that this minimum roughly coincides with the peak of $J_s(T)$ as shown in Fig.5(a). Therefore the second temperature dependent magnetization peak in CaK1144 is actually related to a slower relaxation rate. We notice that the relaxation rates determined by the conventional and dynamical ways have some differences, for example those for $Q$ and $S$ at 6T, but the general trends are similar. In addition, the temperature of the minimum of $Q$ or $S$ may not precisely correspond to that of the maximum $J_s$ (see Fig.5(a)), there is a shift between these two temperatures. This is reasonable since they should have quite different temperature dependence. In Fig.7(b) for BaK122, it is easy to find that $Q$ decreases slowly with increasing temperature. The weak temperature and magnetic field dependence of $Q$ in BaK122 indicate the strong pinning force for vortices in this system. The dropping down of the relaxation rate in BaK122 at high temperatures under some magnetic fields is due to the occurrence of the second peak effect. The relaxation rate determined here is in the same scale of that reported in previous studies\cite{Toshihiro}, while the temperature dependence is a bit different for the data at different magnetic fields. We see a diverging of the relaxation rate at a much higher temperatures under different magnetic fields compared with the previous data. This can be either induced by the different pinning landscape in the two samples or the details of the measurement procedures, such as the measuring time etc., are different.

\begin{figure}
\includegraphics[width=8.5cm]{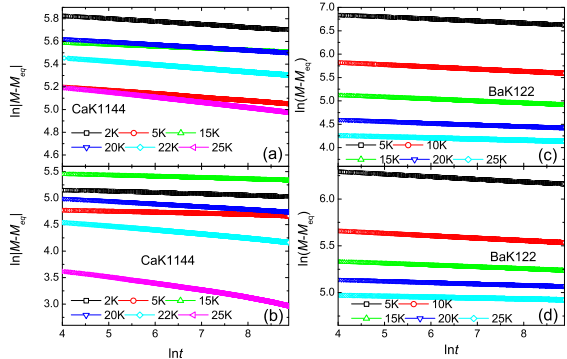}
\caption{(Color online) Magnetization as a function of time on a log-log plot at various temperatures, (a) at 2 T and (b) at 6 T for CaK1144, (c) at 2 T and (d) at 6 T for BaK122.} \label{Fig8}
\end{figure}

In order to further investigate the vortex dynamics in CaK1144 and BaK122, we measured the time decay of magnetization through the method of conventional magnetization relaxation. For each field and temperature, the measuring time was 12000 s for CaK1144, and 7200 s for BaK122. Owing to the long time of this method, we only measured the $M(t)$ for various temperatures at two representative magnetic fields, 2 T and 6 T. But it enables us to compare the vortex dynamics between these two samples. The corresponding $M(t)$ curves in a log-log plot are shown in Fig.~\ref{Fig8}. (a) and (b) for CaK1144 and (c) and (d) for BaK122. It is worth mentioning that, since there is a weak background of the equilibrium magnetization, we subtract a background signal $M_{eq}$ which is determined as the middle value of magnetization measured in the field ascending and descending process. As said before, since the MHLs are quite symmetric, this equilibrium value is actually very small compared to the transient value $M(t)$. Interestingly, in a certain time window, we find a linear relationship between ln$|M-M_{eq}|$ and ln$t$ at low temperatures. In a shorter time scale, the data deviate slightly from this linear relationship. According to Eq.~\ref{eqs}, we know that the slope of curve ln$|M-M_{eq}|$ versus ln$t$ in Fig.~\ref{Fig8} represents the value of relaxation rate $S$. Since the value of $M_{eq}$ is very small, in the theoretical formulas of this paper, we don't write down the $M_{eq}$ and rather use $M$ to represent the corrected magnetization. When temperature warms up to 25 K, the slope increases to a large value for CaK1144. This result suggests clear larger relaxation rates at 25 K in CaK1144. In Fig.~\ref{Fig8} (c) and (d), one can see that the curves at different temperatures are approximately parallel to each other, indicating a similar relaxation rate. The deduced $S$ based on the Eq.~\ref{eqs} has already been presented in Fig.~\ref{Fig7}. It is clear that $S(T)$ shows the similar trend as $Q(T)$. According to the Kim-Anderson model\cite{Anderson}, one can easily see that $M$ is proportional to ln$ t$. Thus, strictly speaking, the vortex motion cannot be described precisely by the linear activation function $U(j)=U_0(1-J_{s}/J_{c})$. It is necessary to incorporate the collective pinning model to explain this behavior.

Taking Eq. (3), (4) and (5) into account, the time dependence of magnetization can be derived as\cite{Yeshurun}
\begin{equation}\label{eqMt}
M(t)=\frac{M_0}{[1+\frac{\mu k_{B}T}{U_{0}}\ln(\frac{t}{t_{0}})]^{1/\mu}}.
\end{equation}
Here $M_{0}$ is the magnetization at the initial measuring time scale, $U_{0}$ and $t_{0}$ are the intrinsic pinning energy and relaxation time, respectively. The timescale of $t_0$ is about $10^{-9} ~ \sim10^{-10}$ seconds. When we adopted this formula to fit $M(t)$ obtained by experiment, we found that it is too hard to get credible $\mu$ values by fitting to a single curve of $M$ versus time, because the time scale is still not long enough and many parameters are involved in the fitting. Thus we use another way, namely the Maley's method, to calculate $\mu$.

\subsection{Scaling based on the Maley's method}

\begin{figure}
\includegraphics[width=8.5cm]{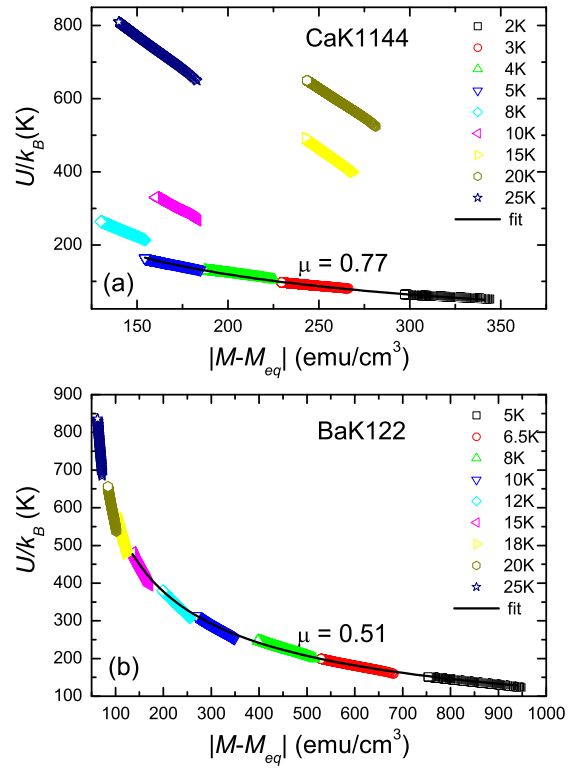}
\caption{(Color online) Activation energy as a function of magnetization derived with Maley's method at various temperatures and magnetic field of 2 T for (a) CaK1144 and (b) BaK122, $M_{eq}$ is the equilibrium magnetization calculated by the averaged value of magnetization measured with increasing and decreasing magnetic fields. The solid lines are the fitting curves with the collective pinning model.} \label{Fig9}
\end{figure}

\begin{figure}
\includegraphics[width=8.5cm]{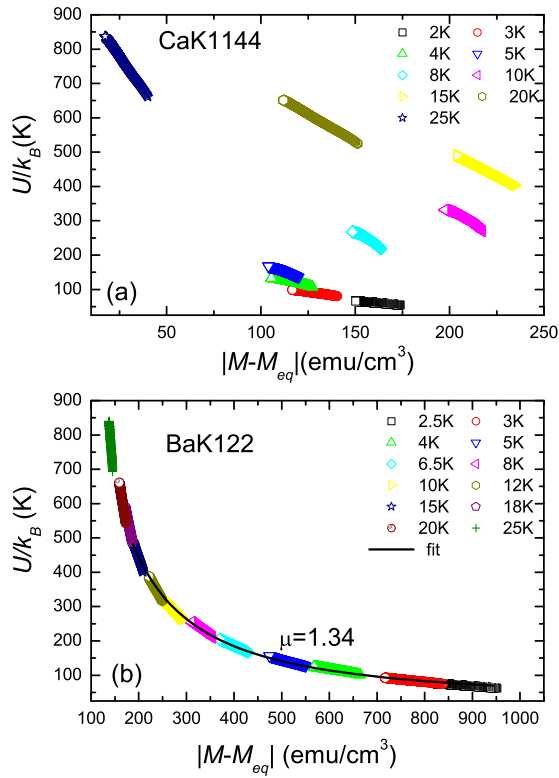}
\caption{(Color online) Activation energy as a function of magnetization determined with Maley's method at various fixed temperatures and a magnetic field of 6 T for (a) CaK1144 and (b) BaK122. The solid line in (b) is the fitting curve with the collective pinning model.} \label{Fig10}
\end{figure}

According to the model of thermally activated flux motion, one can derive the following expression, which was first suggested by Maley et al.\cite{Maley}.
\begin{equation}\label{eqmar}
 U/k_{B}=-T [ln|dM/dt|+ln(B v_0/\pi d)].
\end{equation}
 Here $v_0$ is the attempting moving velocity, $d$ is the dimension of the sample. The second term in the bracket could be regarded as a constant and denoted by $C$ since it is logarithmically related to $Bv_0/\pi d$. Thus Maley et al. proposed that, at a certain magnetic field $B$, the $U(M)$ deduced from $M(t)$ measured at various temperatures should fall onto a smooth curve when the value of $C$ is properly chosen. Once we have a $U(M)$ dependence in a wide regime of $M$ ($M \propto J_s$), and considering the Eq.~\ref{equ}, we are able to get the value of $\mu$. We know that the glassy exponent $\mu$ is a crucial parameter within the collective creep theory. The values of $\mu$ =1/7, 3/2 and 7/9 were predicted to correspond to the cases of vortex motion of single vortex, small bundles and large bundles, respectively\cite{Feigel}. In Fig.~\ref{Fig9} and Fig.~\ref{Fig10}, we present the magnetization dependence of activation energy derived with Maley's method for both samples at 2 T and 6 T, respectively.

 In Fig.~\ref{Fig9}(a), the activation energy calculated by Eq.~\ref{eqmar} for CaK1144 is shown. It is evident that the Maley's method is unapplicable when the temperature is larger than 8 K since the data do not fall onto a single curve. In some cases, we have to scale $U(M)$ curves with $g(T/T_{c})$ function ($g(T/T_{c})=(1-T/T_{c})^{n}$) if the activation energy at high temperatures deviates from one smooth curve\cite{Mchenry}. But we find that it is still unapplicable even this temperature dependent $g(t)$ is considered. In contrast to the case in CaK1144, as shown in Fig.~\ref{Fig9}(b), the $U(M)$ curves for BaK122 at various temperatures are clearly falling onto one smooth curve. This indicates that the Maley's method can work well for BaK122, but cannot for CaK1144. The suitable $C$ derived here for BaK122 equals to 25. In order to know whether the collective vortex creep model is applicable to both samples or not, we use the Eq.~\ref{equ} to fit the resultant $U(M)$ relations. Since the scaling for the CaK1144 sample seems to work for only several temperatures below 8K, we do the fitting only with these very limited data. The value of $\mu$ extracted from the fitting is 0.77 for CaK1144(T $\leq$ 5 K). However the fitting for the BaK122 sample can be done for much wider temperatures, we get $\mu$ = 0.51 for BaK122( T $ \leq$ 18 K) at the magnetic field of 2 T. These $\mu$ values locate in the region of single vortex ($\mu$ = 1/2) to small bundles ($\mu$ = 3/2) cases at 2 T. In Fig.~\ref{Fig10}, the similar behavior of $U(M)$ appears for the data at 6 T. It is obvious that the data of $U(M)$ at various temperatures deviate from one smooth curve for CaK1144. However, $U(M)$ curves for BaK122 at various temperatures seem fall onto one single curve and can be perfectly fitted by the collective pinning model. The corresponding $\mu$ value derived from the fitting is 1.34 which is close to 3/2, indicating that small bundles of vortices dominate in this case. And it should be noticed that the Maley's method is valid in a wide temperature regime ranging from 2.5 K to 20 K for BaK122.

 Now we briefly discuss about the applicability of the Maley's method to these two different systems and the message of vortex pinning. This method is based on the assumption that the temperature dependence of $U(M)$ is insignificant and the initial values of $|M-M_{eq}|$ should decrease with increasing temperature. For BaK122, these two conditions are satisfied so that it is an useful way to determine the correlation of activation energy versus magnetization. On the contrary, for CaK1144, the initial values of $|M-M_{eq}|$ reveal a non-monotonic dependence with increasing temperature, due to the very strange second peak of magnetization versus temperature in this system. Therefore, adjusting the $U(M)$ for data measured in a wide temperature region to a smooth curve is impossible. If we focus only on the $U(M)$ curves at low temperatures for CaK1144, we are still able to get the credible $\mu$ values. However, it is clear that the Maley's method cannot work for the CaK1144 system in a wide temperature region, indicating that the regions of vortex dynamics may change with temperature in CaK1144, this is also the reason for the second peak effect of magnetization versus temperature found for this sample.

\subsection{Explanation for different behaviors of peak effect in CaK1144 and BaK122}
Here we would like to give qualitative explanations for the different behaviors of second peak effect of magnetization in BaK122 and CaK1144. Let us firstly discuss the system BaK122, the second magnetization peak appears with increasing magnetic field, this has been observed in many other systems. One commonly accepted picture is that the vortex dynamics changes from the elastic motion to plastic motion\cite{Salemsuguisecond}. When the magnetic field is enhanced, the vortices become more and more crowded and the elastic energy as well as the shear module $C_{66}$ of the vortex system is increased. This leads to the increase of the vortex pinning force per volume and so does the magnetization. However, to a certain extent, the system is unbearable for a further increase of magnetic field and the interacting energy, some dislocations will be formed and plastic vortex motion starts to appear and gradually dominate the vortex motion\cite{Abulafia}, this will lead to the decrease of the transient critical current density $J_s$ and also the magnetization. As shown in Fig.4(b) and Fig.6(b), we see the anti-correlation between the field dependence of magnetization relaxation rate and magnetization, which gives support to this picture. For the BaK122 system, at a certain magnetic field below the second peak field $H_p$, for example 2 T or 6 T, the same vortex collective pinning regime may work at different temperatures. This allows us to see the successful scaling based on the Maley's method. In a previous study\cite{VandeBeekPRLBaFeAsP}, the magnetization was measured up to 2 Tesla for both the iso-valent doped sample Ba(FeAs$_{0.67}$P$_{0.33})_2$ and other samples including the optimally doped BaK122. The latter were called as the "charged doping" samples which all exhibit the second magnetization peak. The data show that in the iso-valent doped sample Ba(FeAs$_{0.67}$P$_{0.33})_2$ the magnetization shows a monotonic decay with magnetic field, while other samples with "charged doping" show the second peak which can be interpreted as the consequence of the doping induced quasiparticle scattering and thus the collective pinning model applies. The authors argue that the collective pinning induced by the disorders may be the reason for the second peak effect. This picture is certainly interesting. In our CaK1144 samples, we suppose that we do not have this "charged doping" as the quasiparticle scattering centers. This explanation seems to be consistent with our experimental observations.

For the system CaK1144, we do not see a clear second peak effect on the magnetic field dependence of magnetization, but rather on the temperature dependent curve. Naively we would conclude that both systems are hole doped and near the optimal doping point, so that the vortex pinning and vortex dynamics should be similar. However, a closer scrutiny on the atomic structure we see the difference. For BaK122, the element K is randomly doped to the Ba sites, therefore the in-plane strain or in-plane lattice constant are uniform along c-axis. This leads to a rather uniform electronic properties along c-axis. For CaK1144, since the radii of elements K and Ca are quite different, that is why we have a new phase 1144, instead of 122\cite{Akiraiyo}. We can imagine that the actual doping level along c-axis may have a slight alternation leading to a periodic distribution of charge carrier densities of the electronic properties. Moreover, even the doping level may be rather uniform due to the high mobility of the charge carriers, the in-plane strain should inevitably alternates along c-axis due to the different radii of Ca and K atoms. Therefore the vortices will exhibit different features compared with BaK122. At a certain magnetic field, with increasing temperature, due to the c-axis alternation of electronic properties, the vortices are easy to entangle each other. When temperature is increased in the intermediate region, the vortex entanglement will enhance the vortex pinning, leading to an increase of the transient critical current density $J_s$ and the remanent magnetization. While with further increase of temperature, the elastic vortex motion cannot be sustained since the strong entanglement of vortices would lead to the vortex cutting, which produces the dislocations. The motion of these dislocations will lead to the plastic vortex motion, which  will prevail over the elastic motion. The combined effect of elastic energy of strongly entangled vortex bundle and the tilt module $C_{44}$ now plays the key role for the formation of the second peak effect of magnetization versus temperature. As shown in Fig.5(a) and Fig.7(a), we see the anti-correlation between the temperature dependence of magnetization relaxation rate and magnetization, which gives support to this picture. This observation and proposed picture are quite interesting and certainly deserve further investigation.

\begin{figure}
\includegraphics[width=8.5cm]{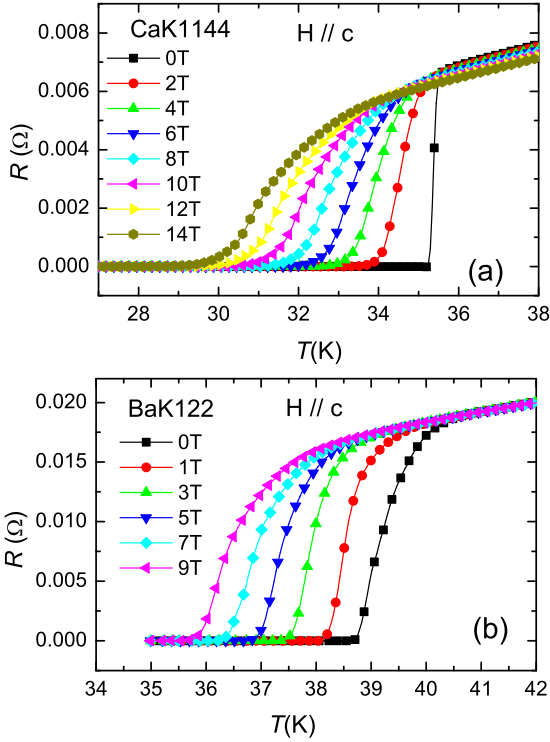}
\caption{(Color online) Temperature dependence of resistivity under different magnetic fields for (a) CaK1144 and (b) BaK122. One can see that the superconducting transitions are quite sharp and the broadening of resistive transition is not very wide for these samples.} \label{Fig11}
\end{figure}

\begin{figure}
\includegraphics[width=8.5cm]{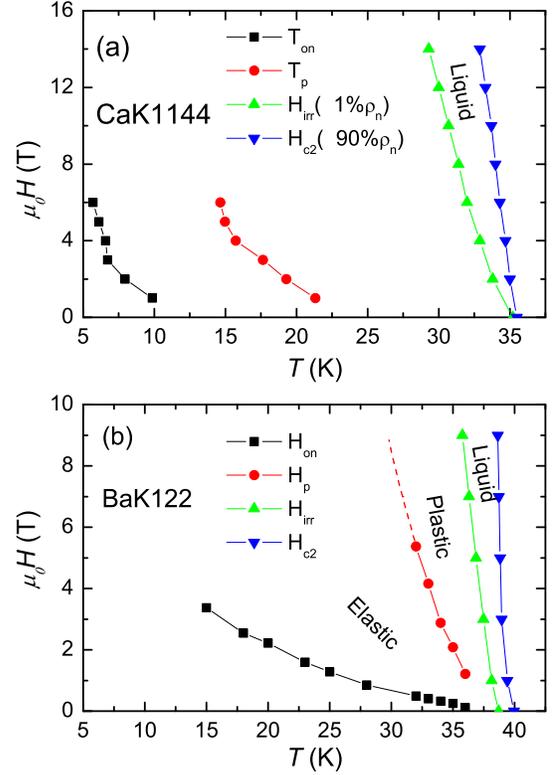}
\caption{(Color online) Different regimes of vortex dynamics on the $H-T$ panel for (a) CaK1144 and (b) BaK122. Here $T_{on}$ and $T_{p}$ are taken from $J(T)$ curves for CaK1144. $H_{on}$ and $H_{p}$ are taken from $M(H)$ curves for BaK122. The terms liquid, plastic and elastic here represent the regimes of vortex liquid, plastic motion and elastic motion, respectively. The dashed line is a guide to the eye.} \label{Fig12}
\end{figure}

\subsection{Different regimes of vortex dynamics on the H-T phase diagram}
In Fig.11 (a) and (b) we show the temperature dependence of resistivity under different magnetic fields for CaK1144 and BaK122, respectively.
One can see that the broadening of resistive transitions under different magnetic fields are rather weak. This indicates that the regime for flux
flow is rather narrow, in contrast with many cuprate systems\cite{Palstra,Pradhan,YoichiAndo}. This may be induced by rather small anisotropy
of $m_c/m_{ab}$ in present iron based superconductors, where $m_c$ and $m_{ab}$ are the mass matrix elements along c-axis and ab-plane.
By the way the samples for the magnetization and resistivity measurements are different, therefore their $T_c$ values can be slightly different. In addition, we can see a slight negative radioresistance feature in the normal state of the CaK1144 sample, which may not be intrinsic, it could be induced by the electrode configuration since the sample has a roughly square shape which cannot allow us to make a standard four probe configuration. By taking the criterions of 90$\%\rho_n$ and 1$\%\rho_n$, we determine the upper critical field $H_{c2}$ and irreversibility field $H_{irr}$ and shown them in Fig.~\ref{Fig12} (a) for CaK1144, and (b)for BaK122. For the CaK1144 system, we define the temperature $T_{on}$ and $T_{p}$ as the minimum and maximum on the $J_s(T)$ curve. Similarly, we define the magnetic fields $H_{on}$ and $H_{p}$ as the the valley and the second peak position on the $M(H)$ curve for BaK122. Although the temperature and magnetic field dependence of magnetization are very different for the two samples, the behavior of $H_{irr}$ and $H_{c2}$ are quite similar for the two systems. The slopes of $H_{irr}$ and $H_{c2}$ for both samples are very steep. Since the irreversibility line usually corresponds to the vortex depinning boundary, we thus can define the region between $H_{irr}$ and $H_{c2}$ as vortex liquid. For BaK122, since in the region above the second magnetization peak, the relaxation rate rises up gradually, we thus define the phase line $H_p(T)$ as the crossover from low temperature elastic to high temperature plastic motions. For the CaK1144, however this second peak occurs on the curve of M versus T, and the crossover field/temperature seem too low. We thus cannot define it as the crossover between the elastic and plastic motions. The vortex liquid regime for flux flow is very small, which suggests a very good potential of practical applications of both systems. Concerning the second peak boundaries of $T_p(H)$ for CaK1144 and $H_p(T)$ for BaK122, one can see that there is big difference. The $T_p(H)$ in CaK1144 locates at much lower values of magnetic fields compared with $H_p(T)$ for BaK122. This clearly shows that they should have very different origins.

\section{Conclusions}
In summary, we have measured the magnetization relaxation and resistivity in two systems CaK1144 and BaK122 with similar superconducting transition temperatures. By using the so-called dynamical and conventional magnetization relaxation methods, we have extensively studied the vortex dynamics in these two systems. Although we find some similarities between them, such as similar critical current density values and the phase boundaries of $H_{irr}$ and $H_{c2}$, clear distinctions are found between them. These include (1) The second peak effect on $M(H)$ or $J_{s}(H)$ curves is present for BaK122, but is absent for CaK1144; however a second peak effect appears on the temperature dependence of critical current density $J_s$ for CaK1144. We give qualitative explanations for different second peak effect in the two different systems. (2) The Maley's method seems not working for CaK1144 in wide temperature regions, but it works perfectly for BaK122. We attribute these differences to the distinct vortex dynamics in the two systems. (3) For BaK122, based on the Maley's method, the $U(M)$ or $U(J_s)$ dependence can be obtained in wide region of current density by using the data measured at different temperatures. A fit to the resultant $U(J_s)$ can lead to the determination of glassy exponent $\mu$ in BaK122. For CaK1144, however, the vortex dynamics varies strongly with temperature, which does not allow us to adopt the Maley's method and use the collective pinning model. Our comparative studies should be helpful to comprehend the understanding of vortex dynamics in pnictide superconductors. Clearly further experiments and studies are still necessary for discerning the fundamental reasons for the different behaviors of vortex dynamics in the two systems.

\section*{ACKNOWLEDGMENTS}
This work was supported by the National Key Research and Development Program of China (Grant Nos. 2016YFA0300401), and the National Natural Science Foundation of China (Grant Nos. A0402/11534005).

$^\dag$ Present address: School of Physics, Sun Yat-Sen University, Guangzhou 510275, China


\begin{thebibliography}{00}

\bibitem{HosonoJACS2008} Y. Kamihara, T. Watanabe, M. Hirano, and H. Hosono, J. Am. Chem. Soc. 130, 11, 3296-3297(2008).
\bibitem{Toshihiro} T. Taen, F. Ohtake, S. Pyon, T. Tamegai, and H. Kitamura, Supercond. Sci. Technol. \textbf{28}, 085003(2015).
\bibitem{Senatore} C. Senatore, R. Flukiger, M. Cantoni, G. Wu, R. H. Liu, and X. H. Chen, Phys. Rev. B \textbf{78}, 054514(2008).
\bibitem{Prozorov} R. Prozorov, M. E. Tillman, E. D. Mun, and P. C. Canfield, New Journal of Physics \textbf{11}, 035004(2009).
\bibitem{HuanYangAPL} H. Yang, H. Q. Luo, Z. S. Wang, and H. H. Wen, Appl. Phys. Lett. \textbf{93}, 142506(2008).
\bibitem{BingShenPRB} B. Shen, P. Cheng, Z. S. Wang, L. Fang, C. Ren, L. Shan, and H. H. Wen, Phys. Rev. B \textbf{81}, 014503(2010).
\bibitem{Anderson} P. W. Anderson, Phys. Rev. Lett. \textbf{9}, 309(1962).
\bibitem{Feigel} M. V. Feigel¡¯man, V. B. Geshkenbein, A. I. Larkin, and V. M. Vinokur, Phys. Rev. Lett. \textbf{63}, 2303(1989).
\bibitem{Fisher} M. P. A. Fisher, Phys. Rev. Lett. \textbf{62}, 1415(1989).
\bibitem{HuanYangPRB} H. Yang, C. Ren, L. Shan, and H. H. Wen, Phys. Rev. B \textbf{78}, 092504(2008).
\bibitem{Haberkorn} N. Haberkorn, M. Miura, B. Maiorov, G. F. Chen, W. Yu, and L. Civale, Phys. Rev. B \textbf{84}, 094522(2011).
\bibitem{YiYu} Y. Yu, C. C. Wang, Q. J. Li, H. Wang, and C. J. Zhang,
    Supercond. Sci. Technol. \textbf{30}, 085007(2017).
\bibitem{ShyamSundarPRB} S. Sundar, S. Salem-Sugui, Jr., H. S. Amorim, H. H. Wen, K. A. Yates, L. F. Cohen, and L. Ghivelder, Phys. Rev. B \textbf{95}, 134509(2017).
\bibitem{Bonura} M. Bonura, E. Giannini, R. Viennois, and C. Senatore, Phys. Rev. B \textbf{85}, 134532(2012).
\bibitem{Das} P. Das, Ajay D. Thakur, Anil K. Yadav, C. V. Tomy, M. R. Lees, G. Balakrishnan, S. Ramakrishnan, and A. K. Grover, Phys. Rev. B \textbf{84}, 214526(2011).
\bibitem{Pramanik} A. K. Pramanik, L. Harnagea, C. Nacke, A. U. B. Wolter, S. Wurmehl, V. Kataev, and B. B¡§uchner, Phys. Rev. B \textbf{83}, 094502(2011).
\bibitem{VanderBeek} C. J. van der Beek, G. Rizza, M. Konczykowski, P. Fertey, I. Monnet, T. Klein, R. Okazaki, M. Ishikado, H. Kito, A. Iyo, H. Eisaki, S. Shamoto, M. E. Tillman, S. L. Bud¡¯ko, P. C. Canfield, T. Shibauchi, and Y. Matsuda, Phys. Rev. B \textbf{81}, 174517(2010).
\bibitem{Nakajima} Y. Nakajima, Y. Tsuchiya, T. Taen, T. Tamegai, S. Okayasu, and M. Sasase, Phys. Rev. B \textbf{80}, 012510(2009).
\bibitem{Shegeyuki} S. Ishida, D. Song, H. Ogino, A. Iyo, H. Eisaki, M. Nakajima, J. Shimoyama, and M. Eisterer, Phys. Rev. B \textbf{95}, 014517(2017).
\bibitem{ZhaoshengWang} Z. S. Wang, H. Q. Luo, C. Ren, and H. H. Wen, Phys. Rev. B \textbf{78}, 140501(R)(2008).
\bibitem{Akiraiyo} A. Iyo, K. Kawashima, T. Kinjo, T. Nishio, S. Ishida, H. Fujihisa, Y. Gotoh, K. Kihou, H. Eisaki, and Y. Yoshida, J. Am. Chem. Soc. 138, 10, 3410-3415(2016).
\bibitem{Meier} W. R. Meier, T. Kong, S. L. Bud'ko, and P. C. Canfield, Phys. Rev. Materials \textbf{1}, 013401(2017).
\bibitem{Luohuiqian} H. Q. Luo, Z. S. Wang, H. Yang, P. Cheng, X. Y. Zhu, H. H. Wen, Supercond. Sci. Technol. \textbf{21}, 125014(2008).
\bibitem{Jirsa1997} M. Jirsa, et al, Supercond. Sci. Technol. \textbf{10}, 484(1997).
\bibitem{Yeshurun} Y. Yeshurun, A. P. Malozemoff, and A. Shaulov, Rev. Mod. Phys. \textbf{68}, 911(1996).
\bibitem{Salemsugui} S. Salem-Sugui, Jr., L. Ghivelder, A. D. Alvarenga, L. F. Cohen, H. Q. Luo, and X. Y. Lu, Phys. Rev. B \textbf{84}, 052510(2011).
\bibitem{Prozorovsecond} R. Prozorov, N. Ni, M. A. Tanatar, V. G. Kogan, R. T. Gordon, C. Martin, E. C. Blomberg, P. Prommapan, J. Q. Yan, S. L. Bud¡¯ko, and P. C. Canfield, Phys. Rev. B \textbf{78}, 224506(2008).
\bibitem{Bean} Charles P. Bean, Rev. Mod. Phys. \textbf{36}, 31(1964).
\bibitem{YangRenSun} Y. R. Sun, J. R. Thompson, H. R. Kerchner, D. K. Christen, M. Paranthaman, and J. Brynestad, Phys. Rev. B \textbf{50}, 3330(1994).
\bibitem{Yang} G. Yang, P. Shang, S. D. Sutton, I. P. Jones, J. S. Abell, and C. E. Gough, Phys. Rev. B \textbf{48}, 4054(1993).
\bibitem{Sun} D. L. Sun, Y. Liu, and C. T. Lin, Phys. Rev. B \textbf{80}, 144515(2009).
\bibitem{Pramaniksecond} A. K. Pramanik, S. Aswartham, A. U. B. Wolter, S. Wurmehl, V. Kataev, and B. B¨¹chner, Journal of Physics: Condensed Matter \textbf{25}, 495701(2013).
\bibitem{Yamamoto} A. Yamamoto, J. Jaroszynski, C. Tarantini, L. Balicas, J. Jiang, A. Gurevich, D. C. Larbalestier, R. Jin, A.,S. Sefat, M. A. McGuire, B. C. Sales, D. K. Christen, and D. Mandrus, Appl. Phys. Lett. \textbf{94}, 062511(2009).
\bibitem{Ahmad} D. Ahmad, W. J. Choi, Y. I. Seo, S-G Jung, Y. C. Kim, S. Salem-Sugui Jr., T. Park, and Y. S. Kwon, Supercond. Sci. Technol. \textbf{30}, 105006(2017).
\bibitem{KonczykowskiPRB2012} M. Konczykowski, C. J. van der Beek, M. A. Tanatar, H. Q. Luo, Z. S. Wang, B. Shen, H. H. Wen, and R. Prozorov, Phys. Rev. B \textbf{86}, 024515(2012).
\bibitem{Vandalen} A. J. J. van Dalen, M. R. Koblischka, R. Griessen, Physica C: Superconductivity and its applications \textbf{259}, 157-167(1996).
\bibitem{YanjingJiao} Y. J. Jiao, W. Cheng, Q. Deng, H. Yang, H. H. Wen, Physica C: Superconductivity and its applications \textbf{545}, 43-49(2018).
\bibitem{Malozemoff} A. P. Malozemoff, Physica C: Superconductivity \textbf{185}, 264-269(1991).
\bibitem{Zeldov} E. Zeldov, N. M. Amer, G. Koren, A. Gupta, R. J. Gambino, and M. W. McElfresh, Phys. Rev. Lett. \textbf{62}, 3093(1989).
\bibitem{WenHH1995} H. H. Wen, H. G. Schnack, R. Griessen, B. Dam, J. Rector, Physica C \textbf{241}, 353-374(1995).
\bibitem{Maley} M. P. Maley, J. O. Willis, H. Lessure, and M. E. McHenry, Phys. Rev. B \textbf{42}, 2639(R)(1990).
\bibitem{Mchenry} M. E. McHenry, S. Simizu, H. Lessure, M. P. Maley, J. Y. Coulter, I. Tanaka, and H. Kojima, Phys. Rev. B \textbf{44}, 7614(1991).
\bibitem{Salemsuguisecond} S. Salem-Sugui, Jr., L. Ghivelder, A. D. Alvarenga, L. F. Cohen, K. A. Yates, K. Morrison, J. L. Pimentel, Jr., H. Q. Luo, Z. S. Wang, and H. H. Wen, Phys. Rev. B \textbf{82}, 054513(2010).
\bibitem{VandeBeekPRLBaFeAsP}C. J. van der Beek, M. Konczykowski, S. Kasahara, T. Terashima, R. Okazaki,T. Shibauchi, and Y. MatsudaK. Phys. Rev. Lett. \textbf{105}, 267002(2010).
\bibitem{Abulafia} Y. Abulafia, A. Shaulov, Y. Wolfus, R. Prozorov, L. Burlachkov, Y. Yeshurun, D. Majer, E. Zeldov, H. W¨¹hl, V. B. Geshkenbein, and V. M. Vinokur, Phys. Rev. Lett. \textbf{77}, 1596(1996).
\bibitem{Palstra} T. T. M. Palstra, B. Batlogg, R. B. van Dover, L. F. Schneemeyer, and J. V. Waszczak, Phys. Rev. B \textbf{41}, 6621(1990).
\bibitem{Pradhan} A. K. Pradhan, M. Muralidhar, Y. Feng, M. Murakami, K. Nakao, and N. Koshizuka, Phys. Rev. B \textbf{64}, 172505(2001).
\bibitem{YoichiAndo} Y. Ando, G. S. Boebinger, A. Passner, L. F. Schneemeyer, T. Kimura, M. Okuya, S. Watauchi, J. Shimoyama, K. Kishio, K. Tamasaku, N. Ichikawa, and S. Uchida, Phys. Rev. B \textbf{60}, 12475(1999).









\end{thebibliography}
\end{document}